\documentclass[10pt]{article}
\usepackage[OE]{style}
\usepackage{varioref}
\usepackage{csquotes}
\begin{document}

\title{Dynamic defects in photonic Floquet topological insulators}

\author{Christina J\"org,\authormark{1,*}, Fabian Letscher,\authormark{1,2}, Michael Fleischhauer,\authormark{1} and Georg von Freymann\authormark{1,3}}

\address{\authormark{1}Physics Department and Research Center OPTIMAS, University of Kaiserslautern,  67663 Kaiserslautern, Germany\\
\authormark{2}Graduate School Materials Science in Mainz, 67663 Kaiserslautern, Germany\\
\authormark{3}Fraunhofer Institute for Industrial Mathematics ITWM, 67663 Kaiserslautern, Germany}

\email{\authormark{*}cjoerg@physik.uni-kl.de}


\begin{abstract}
Edge modes in topological insulators are known to be robust against defects. We investigate if this also holds true when the defect is not static, but varies in time.
We study the influence of defects with time-dependent coupling on the robustness of the transport along the edge in a Floquet system of helically curved waveguides.
Waveguide arrays are fabricated via direct laser writing in a negative tone photoresist. We find that single dynamic defects do not destroy the chiral edge current, 
even when the temporal modulation is strong.
Quantitative numerical simulation of the intensity in the bulk and edge waveguides confirms our observation.
\end{abstract}

\ocis{Array waveguide devices; Microstructure fabrication.} 

\section{Introduction}
Topological insulators are materials that are insulators in their volume but conduct current at their surfaces. This surface current is robust against certain defects \cite{asboth2016}. In this paper we study if the surface current is also robust against dynamic defects, i.e. imperfections at the surface that show a time-dependent behavior.

The research on topological insulators goes back to the discovery of the Quantum Hall effect \cite{Klitzing}:
Applying a high magnetic field to a two dimensional electron gas at low temperatures results in a drop of the resistance at certain field strengths. This means that current is conducted almost dissipation-less. 
Although by now many semiconductors have been proven to show topological behavior, engineering topological insulators in semiconductor compounds often requires changing the material properties, e.g. by doping \cite{BiTe}.
A different approach is to apply a time-periodic drive. It has been shown that such a time-periodic drive can induce topological behavior in a system, which is topologically trivial without the drive\cite{konig_quantum_2007,Lindner}. Such a system is called a Floquet topological insulator (FTI). The big advantage of this method is that the properties of the system can be tuned externally. 
Floquet periodic drive can also be used to create effective magnetic fields in model systems based on cold atoms or photons which scarcely couple to real magnetic fields.

In FTIs defects are generically time-dependent due to the time-periodicity of the driving field. This makes them more complex than defects in static systems. For example it has been proposed that disorder can induce transitions between topologically trivial and non-trivial systems in FTIs \cite{Disorder}. This raises the question if a topological edge state in an FTI is still robust in the presence of a time-dependent defect.

In solids, defects are not easy to control, and systematic tuning of the parameters is difficult. Therefore, many model systems have been proposed, in which specific effects can be studied methodically. Among them are experiments with ultracold atoms \cite{Aidelsburger,Esslinger}, optical ring resonators \cite{hafezi2},  gyromagnetic photonic crystals \cite{wang} and optical waveguide arrays \cite{Szameit, Diebel, Longhi, Mukherjee}.
In 2013 the first topological insulator for light in the visible spectrum was realized \cite{Szameit}. This made photonic topological insulators not only interesting as model systems, but also for optics itself, e.g. to enable robust optical data transfer \cite{hafezi_delay}.
The setup in \cite{Szameit} consists of an array of evanescently coupled waveguides arranged on a honeycomb lattice.
In waveguide systems the propagation of light is described by the paraxial Helmholtz equation \cite{Szameit}
\begin{align}
\mathrm{i} {\partial_z} {\Psi(x,y,z)} = -\frac{1}{2 k_{\mathrm{wg}}} \left[({\partial_x}^2 + {\partial_y}^2) + V(x,y) \right]\Psi(x,y,z),
\label{eq:Helmholtz}
\end{align}
which is mathematically analog to the Schr\"odinger equation, with an effective potential
\begin{align}
V(x,y)=k_{\mathrm{wg}}^2 \frac{n^2(x,y)-n_{\mathrm{wg}}^2}{n_{\mathrm{wg}}^2}
\label{eq:potential}
\end{align}
(see \cite{Szameit}).
Here, $n(x,y)$ is the refractive index profile and $n_{\mathrm{wg}}$ the refractive index in the waveguide. $k_{\mathrm{wg}}=n_{\mathrm{wg}} 2\pi/\lambda$ is the wavevector in the waveguide and corresponds to the mass in the Schr\"odinger equation, and $\Psi$ is the amplitude of the electric field. Propagation distance along the waveguide axis $z$ corresponds to time. For this reason we can use $z$ and time synonymously.
The intensity distribution of light coupling between waveguides therefore represents the density distribution of electrons in an atomic lattice. Both can be described by the same Hamiltonian in tight binding approximation
\begin{align}
H = c \sum_{m,n} \hat{a}_{m,n}^+ \hat{b}_{m,n} + \hat{a}_{m+1,n}^+\hat{b}_{m,n} + \hat{a}_{m,n+1}^+\hat{b}_{m,n} + \mathrm{h.c.}
\label{eq:H_graphene}
\end{align}
for a honeycomb lattice with two sites a and b per unit cell. 
$\hat{a}_{m+1,n}^+\hat{b}_{m,n}$ creates a particle at site $a_{m+1,n}$ and destroys one at site $b_{m,n}$ with hopping amplitude $c$, where $m,n$ enumerate the sites.
The underlying lattice of the waveguide positions determines the band structure just as the atomic lattice does.
This model system of evanescently coupled waveguides is especially suited to examine edge transport behavior, since a sharp edge is needed for that. In, e.g. cold atoms experiments a sharp boundary is hard to obtain as the atoms are usually trapped in soft potentials. Furthermore, we can insert defects in a controlled way, which is not easily possible in condensed matter systems.

By Fourier transformation of $\hat{a}$ and $\hat{b}$ one obtains the band structure of the system,
which up to this point corresponds to a topologically trivial photonic graphene band structure\cite{plotnik_observation_2014}.
To induce non-trivial topological behavior time reversal symmetry must be broken by applying a periodic drive (Floquet \cite{Aoki,Demler,PhotoinducedQHE,Lindner}). In waveguide systems this is done by curving the waveguides helically \cite{Szameit}. Then, $n(x,y)$ and, therefore, $V$ in equation (\ref{eq:potential}) are no longer stationary, but vary along $z$. To remove this $z$-(time-)dependency, the coordinates are transformed to a reference frame that is rotating, in which the waveguides appear stationary. The effect of the curling then is absorbed by a time-dependent vector potential in the transverse plane
\begin{align}
\boldsymbol{A}(z) =  \frac{2\pi}{Z} k_{\mathrm{wg}} R 
\begin{pmatrix} \sin(2\pi z/Z) , -\cos(2\pi z/Z)\end{pmatrix}
\label{eq:vectorpot}
\end{align}
with $R$ being the helix radius and $Z$ the helix pitch.
$\boldsymbol{A}(z)$ corresponds to an ac-field \cite{Fang}.
Due to the vector potential the coupling $c$ 
acquires an additional phase, the Peierls phase \cite{Peierls}. Thus the complex coupling attains the form
\begin{align}
c \rightarrow c \exp\left(\mathrm{i}\boldsymbol{A}(z)\cdot \boldsymbol{d}\right),
\label{eq:Peierls}
\end{align}
where $\boldsymbol{d}$ is the distance between waveguide sites. 
We look at the time evolution of the waveguide system stroboscopically. The evolution operator $U(Z)$ describes how the light intensity evolves in the coupled waveguide system within a period $Z$. This can be used to define an effective Hamiltonian 
$\hat{\mathcal{H}}_{\mathrm{eff}}$ setting
\begin{align}
U(Z) = \exp \left( -\mathrm{i}  \hat{\mathcal{H}}_{\mathrm{eff}}Z\right).
\label{eq:evolution}
\end{align}
Now, we use the static effective Hamiltonian to characterize the topology of the system. As shown in \cite{Esslinger} the effective Hamiltonian resembles a Haldane phase \cite{Haldane} in the regime of a fast periodic drive. The quasi energy band structure of the waveguide system is shown in Figure \ref{fig:bandstructure} (a) \cite{Demler,PhotoinducedQHE,Aoki,Lindner}. The band structure hosts chiral edge states similar to a Quantum Hall system.
Such an edge state encircles the structure clock- or anti-clock-wise. It cannot reverse its direction due to topological order. This is why these so called Chern insulators are robust against certain defects, in the sense that at a simple defect no backscattering of the edge mode or scattering into the bulk occurs.
\begin{figure}[t]
  \centering
  \def\svgwidth{1\columnwidth} 
\executeiffilenewer{svg/Defekttypen7.svg}{Defekttypen7.pdf}%
{/usr/bin/inkscape -z -D --file="svg/Defekttypen7.svg" --export-pdf="Defekttypen7.pdf" --export-latex}
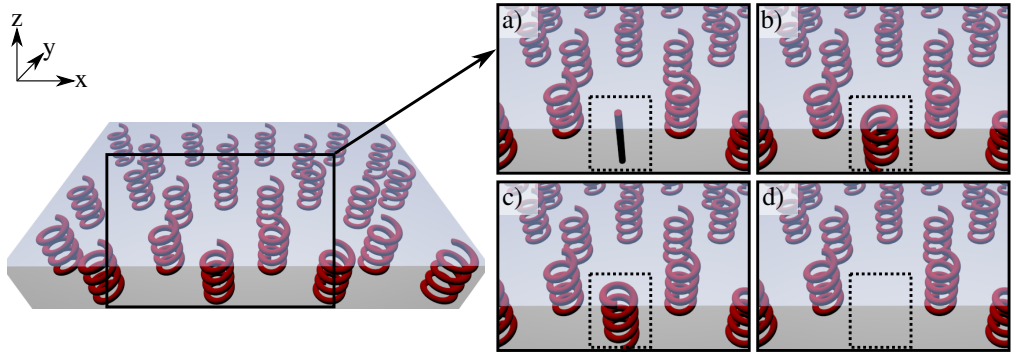%

  \caption{Stylized waveguide samples with different kinds of defects: a) straight defect, b) defect with opposite helicity, c) defect shifted by half a helix pitch, d) missing waveguide for comparison. In each case the defect waveguide substitutes an outer waveguide at the zigzag edge of the honeycomb lattice.}
  \label{fig:defects}
\end{figure}
Defects in a FTI are generically time periodic with the same period as the external drive. Thus we want to examine the robustness of topological edge states against dynamic defects in a waveguide setup of a FTI. 
This setup is suited well, as it comes with almost arbitrary time-resolution.
So far it has been shown in experiments that a topological edge mode survives ``wrong'' edge termination, a missing waveguide \cite{Szameit}, obstacles \cite{wang} and certain kinds of disorder \cite{hafezi2}. However, none of these defects change in time. 
Still, there are experiments that implement time-modulation of the coupling: In \cite{leykam_anomalous_2016, AFI} it is used to realize a photonic anomalous Floquet topological insulator, and in \cite{Non-Hermitian} to induce losses. Yet, both experiments apply this modulation globally for different purposes and do not study the impact on the robustness of the chiral current.
For clarity of the effects, we look at a single defect at the zigzag edge that is curled differently than the other waveguides, thus involving time-dependent coupling. We examine three kinds of dynamic defects (see Figure \ref{fig:defects}); in each case the defect waveguide substitutes an outer waveguide at the zigzag edge of the honeycomb lattice: 
a) a straight waveguide, b) a waveguide with opposite helicity, and c) a waveguide with the same helicity but shifted by half a helix pitch in the $z$-direction (rotation phase-shifted by $\pi$)\footnote{The three examined defects were chosen to share the $Z$-periodicity with the usual waveguides. Using defects with a different $Z$ or even non-periodic trajectory would also be possible.}. 
When transforming the coordinates of the whole system to the rotating frame, the time-dependency cancels for the coordinates of all waveguides but the defect. This is why we call the defect dynamic, in contrast to the rest of the waveguides resting in the rotating frame.
The time-dependency of the defect position results in a time-dependent distance $\boldsymbol{d}(z)$ between defect and neighboring waveguides. This has two effects: First, the real part of the coupling $c$ now becomes time-dependent, as it decreases exponentially with $|\boldsymbol{d}|$ \cite{szameit_hop}. Second, with $\boldsymbol{d}(z)$ also the Peierls phase changes (see equation \ref{eq:Peierls}). Therefore, this might be similar to realizing a magnetic defect in a Quantum Hall phase.

\section{Methods} \label{methods}

\subsection{Sample fabrication} \label{Sample fabrication}
\begin{figure}[ht]
  \centering
  \def\svgwidth{1\columnwidth} 
\executeiffilenewer{svg/Herstellung5.svg}{Herstellung5.pdf}%
{/usr/bin/inkscape -z -D --file="svg/Herstellung5.svg" --export-pdf="Herstellung5.pdf" --export-latex}
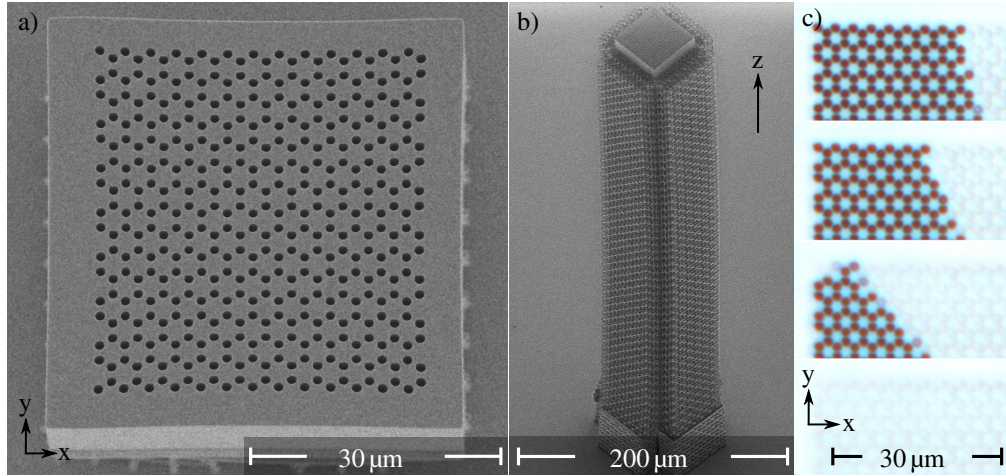%

  \caption{a) and b): Scanning electron microscopy images of inverse waveguide samples: a) top view and b) view tilted by \SI{42}{\degree} (structure with supporting framework). c) The unfilled waveguides (dark) are infiltrated with SU8 by capillary forces. Microscope images are taken at intervals of a few seconds.}
  \label{fig:fabrication}
\end{figure}
To model topological insulators with defects by means of classical optics, we fabricate arrays of evanescently coupled waveguides. 
These waveguides are about \SI{1}{\um} in diameter at an aspect ratio of 1:500, and helically curved.
We fabricate the inverse of the sample by 3D-lithography (direct laser writing, DLW)\cite{hohmann_dlw}. DLW works by two-photon-polymerization of a liquid negative tone photoresist (IP-Dip, Nanoscribe). Aberrations of the laser focus used for writing are corrected by a spatial light modulator \cite{Waller:12, Hering} (for more details of this fabrication method read \cite{hohmann_dlw}).
In standard writing configuration the height of the structures is limited by the working distance of the objective. As the waveguide structures are required to be quite high (about \SI{500}{\um} normal to the substrate), we use DLW in Dip-In configuration \cite{dip-in}, which means that the writing-objective is dipped right into the resist applied to the bottom of a glass cover-sheet. The structure then is built layer by layer (starting at the glass sheet) by moving the objective in the $z$-direction. To minimize the stress onto the structure that occurs due to shrinkage during development, the structure is put onto a grid \cite{grid, Renner}. Besides leading to uniform shrinking of the structure, it also helps to remove the unpolymerized resist from the waveguide canals during development.
The inverse sample is developed in PGMEA and Isopropanol for about $\SI{45}{minutes}$ and $\SI{30}{minutes}$ respectively. Subsequently, the channels are infiltrated with a different material, creating low-loss 3D waveguides (Figure \ref{fig:fabrication}). As infiltration material we use SU8-2 (Microchem). The infiltrated sample is baked on a hotplate to solidify the SU8. The resulting refractive indices are about 1.59 for the SU8 waveguide and 1.54 for the surrounding material (IP-Dip).

The common method to fabricate waveguide arrays is the femtosecond laser writing method \cite{Szameit}: Femtosecond laser pulses locally change the refractive index in a \SI{10}{cm} long glass block by about $\Delta n = 6\cdot10^{-4}$ to $10^{-3}$ \cite{PhDTutorial}. By moving the glass relatively to the laser focus almost arbitrary trajectories of the waveguides can be written. However, the focus determines the cross-section of the waveguides, making them elliptical. Thus, the coupling between waveguides is not isotropic and has to be corrected by adjusting the spacing between waveguides.

Our fabrication method results in circular cross-sections of the waveguide channels. By choosing a higher refractive index contrast of $\Delta n=0.05$ the bending losses can be reduced and tighter curling is possible. At the same time, coupling between waveguides can still be kept large (about one hop per \SI{60}{\um} propagation) by decreasing the spacing between waveguides to about \SI{1.5}{\um}. This also allows to reduce the overall length of the sample to about half a millimeter. Furthermore, the waveguide diameter is chosen small enough to still be in the single-mode regime.
A further degree of freedom is the choice of the infiltration material, allowing to easily tune the refractive index contrast between waveguide and surrounding.

We fabricate two sets of samples for each of the four defect cases with different parameters. The first one is for weak modulation of the coupling between defect and neighbors, i.e. large spacing between waveguides $a=\SI{1.65}{\um}$ and small helix radius $R=\SI{0.36}{\um}$ at a helix pitch of $Z=\SI{72}{\um}$ and waveguide radius of $r=\SI{0.37}{\um}$. 
The coupling constant between two regular waveguides for this set is numerically calculated to be $c\approx\SI{16300}{m^{-1}}$.
Note, that while the distance $\boldsymbol{d}$ between defect and neighboring waveguide varies symmetrically as, e.g. 
\begin{align}
{\boldsymbol{d}=\begin{pmatrix} -a/2+R\cos(2\pi z/Z) , \sqrt{3}/2a+R\sin(2\pi z/Z)\end{pmatrix}}
\label{eq:dist_straight}
\end{align}
for the straight defect, coupling does not, as it decays exponentially with $|\boldsymbol{d}|$ \cite{szameit_hop}.
\mbox{Time-($z$-)}averaging yields average defect coupling constants of only $\pm3\%$ compared to the coupling between two usual waveguides. Therefore we call the defects of this set of parameters weak dynamic defects.

The second set gives strong modulation in coupling, as $a=\SI{1.40}{\um}$ and $R=\SI{0.89}{\um}$ at $Z=\SI{85}{\um}$ and $r=\SI{0.49}{\um}$. Here, the averaged defect coupling constant $c_d$ differs more from the usual coupling $c$ ($c\approx\SI{23000}{m^{-1}}$). For the defect with opposite helicity the defect coupling $c_d$ is about $70\%$ of $c$, for the phase-shifted defect $150\%$ of $c$ and for the straight defect $180\%$ of $c$.
Note that for this set of parameter the straight and the phase-shifted defect are overlapping with their neighbors at certain times.

\subsection{Measurement setup} \label{Measurement setup}
\begin{figure}[ht]
  \centering
  \def\svgwidth{1\columnwidth}
\executeiffilenewer{svg/messaufbau3.svg}{messaufbau3.pdf}%
{/usr/bin/inkscape -z -D --file="svg/messaufbau3.svg" --export-pdf="messaufbau3.pdf" --export-latex}
\input{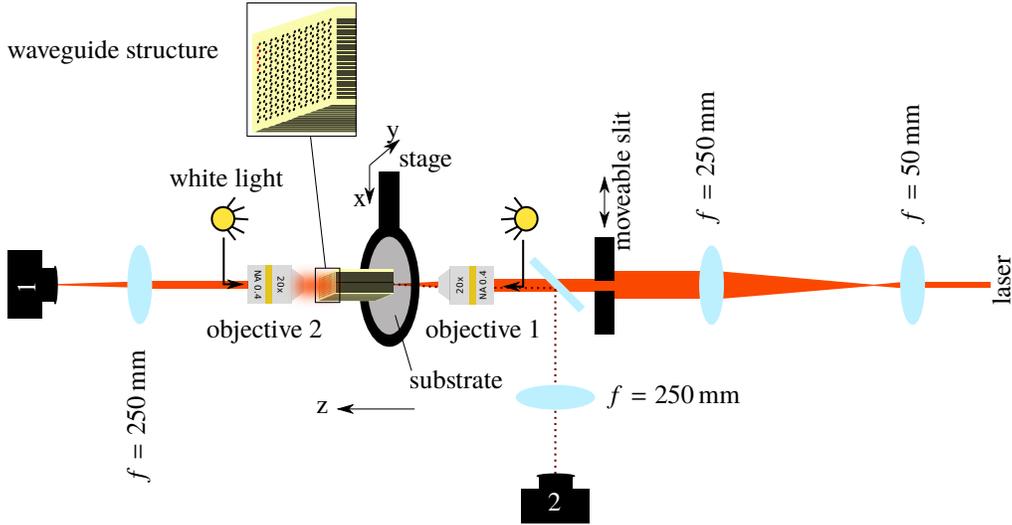}%

  \caption{Schematic illustration of the measurement setup.}
  \label{fig:messaufbau}
\end{figure}
\begin{figure}[ht]
  \centering
  \def\svgwidth{1\columnwidth} 
\executeiffilenewer{svg/VglExpOpti3.svg}{VglExpOpti3.pdf}%
{/usr/bin/inkscape -z -D --file="svg/VglExpOpti3.svg" --export-pdf="VglExpOpti3.pdf" --export-latex}
\input{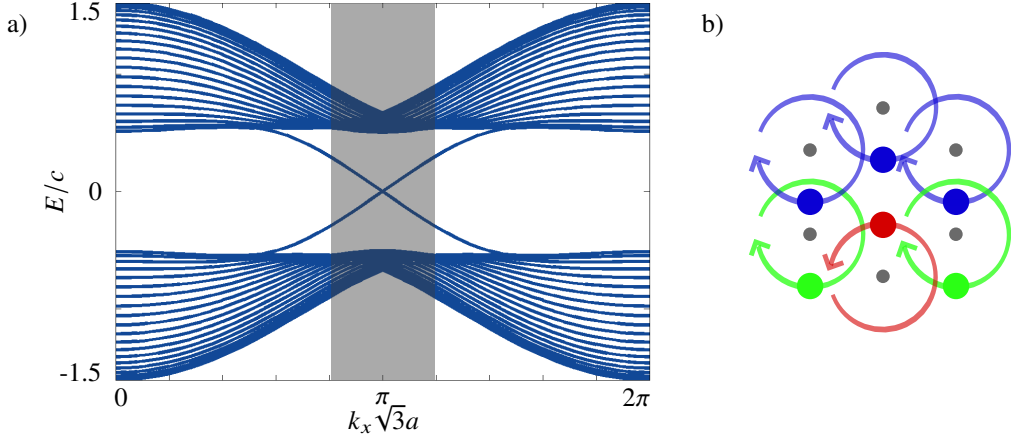}%
  
  \caption{a) Floquet band structure of the honeycomb geometry without defect (quasienergy $E$ divided by coupling $c$). While we assume periodic boundary conditions along $x$, we consider a finite size system along $y$. Therefore, the band structure shows two chiral edge modes on opposite sites in $y$-direction, which are the edge states for the zigzag edge. b) Geometry of defect with opposite helicity for small $a$; green: nearest-neighbors, blue: next-nearest neighbors.}
  \label{fig:bandstructure}
\end{figure}
The beam from a tunable fs-pulsed TiSa-laser ($\SI{680}{nm}$ -- $\SI{1080}{nm}$) is expanded and focused through objective 1 onto the input facet of the waveguide sample (Figure \ref{fig:messaufbau}). The intensity distribution at the output facet is imaged by objective 2 and a lens onto CMOS-camera 1 (Thorlabs). A movable slit in front of objective 1 allows to create an angle in the incoupling beam and thus to select $k_x$-components for excitation (see band structure in Figure \ref{fig:bandstructure}).
To see the excitation area, the reflection of the laser beam at the input facet is imaged by means of a beam splitter, objective 1 and a lens onto CMOS-camera 2. 
By using white light from a lamp we can image the output- (input-) facet of the sample onto camera 1 (camera 2) to identify the waveguide sites.

\section{Results} \label{results}
\subsection{Experiment} \label{Experiment}
\begin{figure}[h]
  \centering
  \def\svgwidth{1\columnwidth} 
\executeiffilenewer{svg/Results7_parula_255_strong.svg}{Results7_parula_255_strong.pdf}%
{/usr/bin/inkscape -z -D --file="svg/Results7_parula_255_strong.svg" --export-pdf="Results7_parula_255_strong.pdf" --export-latex}
\input{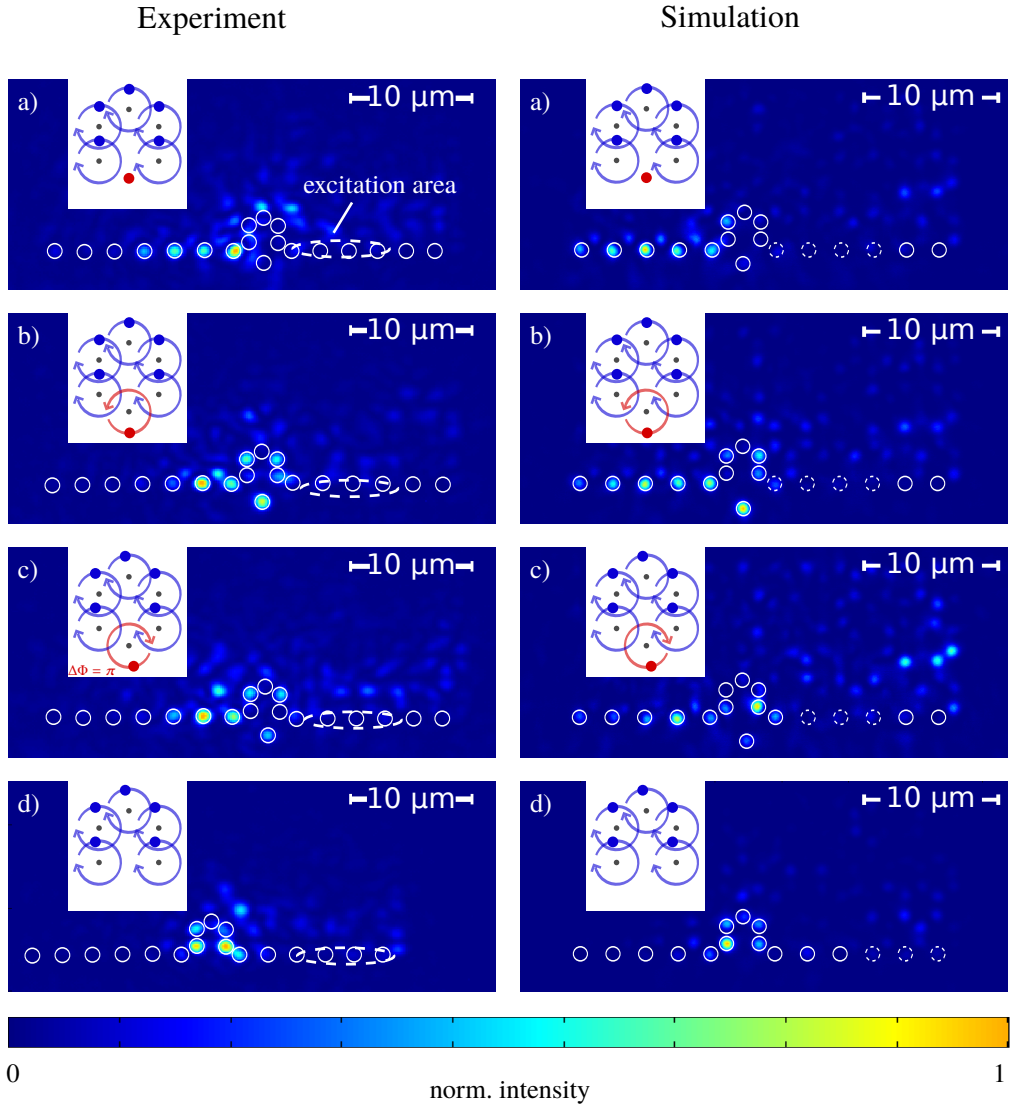}%
  
  \caption{Strong dynamic defects: measured (left) and numeric (right, see chapter \ref{Numerics}) intensity distributions at the output facet of the samples. The edge modes, excited at the location of the dashed ellipse, move around different kinds of defects: a) straight defect, b) defect with opposite helicity, c) defect shifted by half a helix pitch, d) missing waveguide. White circles indicate edge waveguides. Intensity is scaled to maximum separately for each image. The insets show the geometry at the defect. Output after $z=\SI{465(15)}{\um}$ (experiment) and $z=\SI{467.5}{\um}$ (simulation) of propagation. Differences between experiment and simulation are due to fabrication imperfections and slight deviations in the location of excitation.}
  \label{fig:results_strong}
\end{figure}

To observe the effect of a dynamic defect, an edge mode is excited at the position indicated by a dashed ellipse in Figure \ref{fig:results_strong}. This is done by coupling an elliptically shaped laser beam into four waveguides at the zigzag edge. We select $k_x\sqrt{3}a\approx \pi$, where a chiral edge state exists at the zigzag edge (see Figure \ref{fig:bandstructure}).
The excitation area is imaged onto camera 2 and the output distribution onto camera 1. The sample can be moved along the $x$- and $y$-axis by linear actuators to excite different waveguides.
Comparing the location of the excited waveguides at the input and output plane we can see how far and in which direction the edge state has moved along the sample. 
By tuning the wavelength of the laser beam we can tune the coupling constant to some extent \cite{szameit_hop} which has the same effect as fabricating a new sample with different waveguide spacing $a$. We need the edge mode to move far enough along the zigzag edge to see if it walks around the defect. Thus we use a wavelength of $\lambda=\SI{810}{nm}$ on the sample with bigger $a$ (set of weak dynamic defects) and of $\lambda=\SI{710}{nm}$ on the one with smaller $a$ (set of strong dynamic defects).

\begin{figure}[h]
  \centering
  \def\svgwidth{1\columnwidth} 
\executeiffilenewer{svg/Results7_parula_255_weak.svg}{Results7_parula_255_weak.pdf}%
{/usr/bin/inkscape -z -D --file="svg/Results7_parula_255_weak.svg" --export-pdf="Results7_parula_255_weak.pdf" --export-latex}
\input{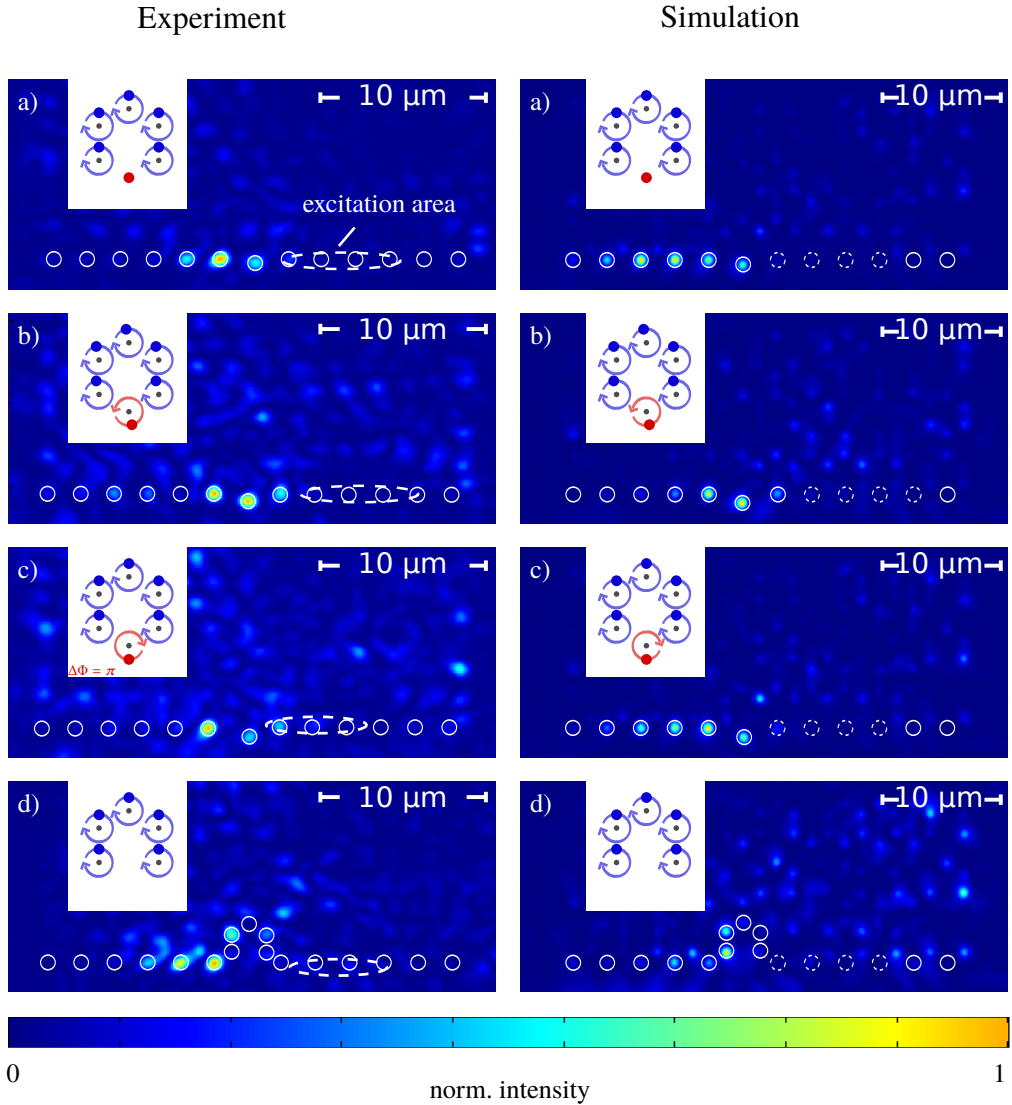}%
  
  \caption{Weak dynamic defects: measured (left) and numeric (right, see chapter \ref{Numerics}) intensity distributions at the output facet of the samples. The edge modes, excited at the location of the dashed ellipse, move around different kinds of defects: a) straight defect, b) defect with opposite helicity, c) defect shifted by half a helix pitch, d) missing waveguide. White circles indicate edge waveguides. Intensity is scaled to maximum separately for each image. The insets show the geometry at the defect. Output after $z=\SI{471(38)}{\um}$ (experiment) and $z=\SI{468}{\um}$ (simulation) of propagation. Differences between experiment and simulation are due to fabrication imperfections and slight deviations in the location of excitation.}
  \label{fig:results_weak}
\end{figure}

Figure \ref{fig:results_weak} shows the intensity distribution in the output plane of the sample for weak defects (small helices) and Figure \ref{fig:results_strong} for strong defects (large helix radius).
Sample heights are $6.5Z$ and $5.5Z$ respectively, and correspond to the propagation distance.
As the heights are not integer multiples of $Z$, the defect seems to be farther away from the edge than the rest. However, the defect's mean position still coincides with a lattice site (see insets to Figure \ref{fig:results_weak} and \ref{fig:results_strong}).

We first look at the case of one waveguide missing at the edge (Figure \ref{fig:results_weak} (d) and \ref{fig:results_strong} (d)). For both sets of parameters the edge state behaves as shown in other setups before, e.g. in \cite{Szameit}: The edge mode moves around the defect, i.e. along the new edge, without observable scattering into the bulk. This indicates that we have indeed a photonic topological insulator. 

While the edge state in a topological insulator has no other possibility than to move around a missing waveguide, its behavior at dynamic defects is yet unknown. The wrong Peierls phase of the defect coupling could lead to scattering into the bulk. However, in all the cases of dynamic defects studied here, the edge mode moves along the edge regardless of the defect (Figure \ref{fig:results_weak} and \ref{fig:results_strong} (a)-(c)). 
Yet, the light does not just move around the defect as in the missing waveguide case, but also through it. 
For large waveguide spacing $a$ (weak defects) the light mainly goes through the defect waveguide. As there is almost no difference between the average defect coupling constant and the usual coupling constant (see section \ref{Sample fabrication}), the defect waveguide seems not to be noted as defect at all, despite the wrong Peierls phase in the coupling.

For small values of $a$ (strong defects) light partially moves around the defect and partially through it. 
At certain times, coupling of the light to non-neighboring waveguides outweighs nearest-neighbor coupling (Figure \ref{fig:bandstructure}). In this case the light partially moves around the defect and partially through it.
This means that time-dependency may also give rise to other effects \cite{leykam_anomalous_2016}.
What is noted though, is that the intensity decreases dramatically in the case of overlapping waveguides. This is attributed to losses due to mode-mismatching, as the cross-section of the joined waveguides is not circular any more. However, this does not contradict the robustness of the edge mode, as the topological protection is only valid against backscattering and not against particle loss \cite{plasmon}.

The experiment suggests that edge modes in topological insulators are still robust in the presence of a single dynamic defect. 

\clearpage

\subsection{Numerical simulations} \label{Numerics}
To analyze the robustness of the transport along the edge quantitatively, numerical calculations are performed.
We examine the portion of intensity in the bulk and edge waveguides along $z$, to see if the defect causes light to scatter into the bulk.
In contrast to the experimental realization, the numerical simulations allow the intensity distribution to be analyzed at multiple values of $z$ in one run. Experimentally, one would need to fabricate many samples of different heights to access multiple $z$-slices. As the numerical calculations match the measurements well (compare Figure \ref{fig:results_weak} and \ref{fig:results_strong}), we use the OptiBPM software (Optiwave), which relies on the beam propagation method.

\begin{figure}[h]
  \centering
  \def\svgwidth{1\columnwidth} 
\executeiffilenewer{svg/weak3.svg}{weak3.pdf}%
{/usr/bin/inkscape -z -D --file="svg/weak3.svg" --export-pdf="weak3.pdf" --export-latex}
\input{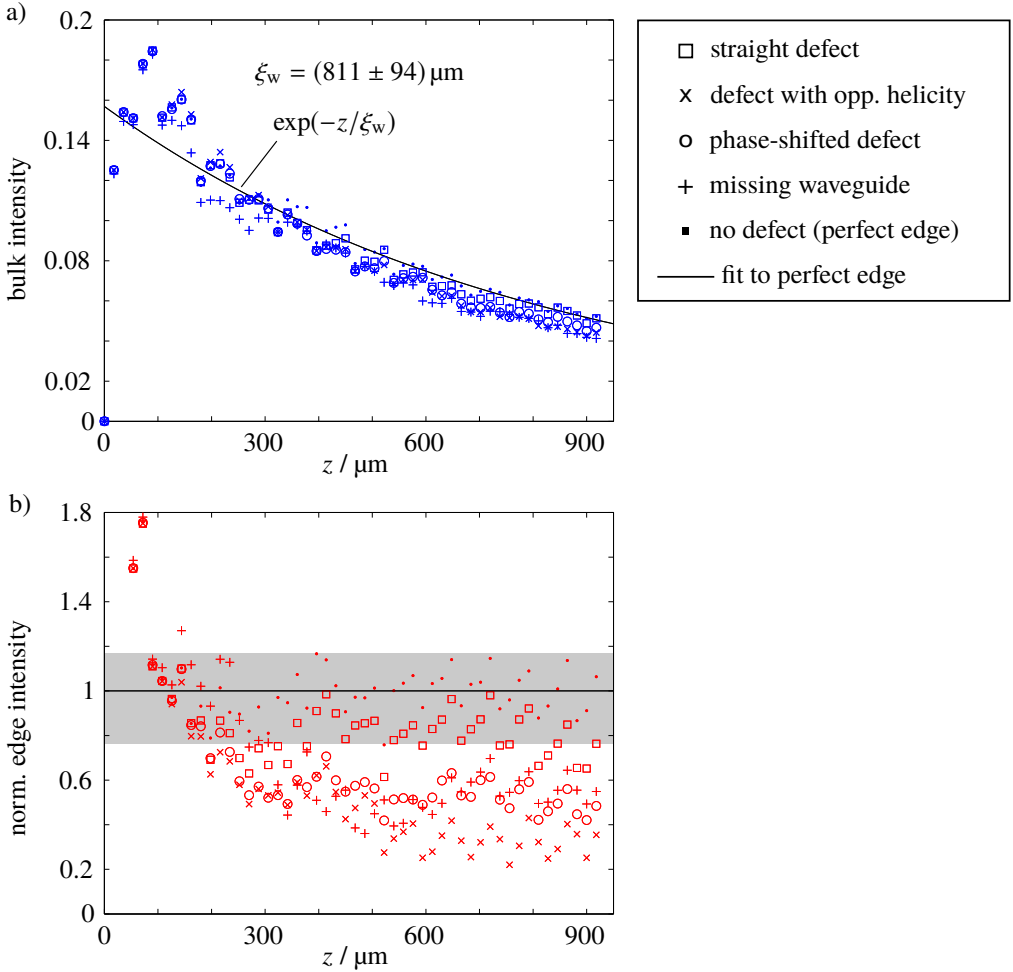}%
  
  \caption{Numerical calculations for weak dynamic defects: Intensity in edge (red) and bulk (blue) for samples with and without defects. b) normalized to fitting curve for the edge intensity of the defect-free sample to remove overall losses.
  }
  \label{fig:opti_weak}
\end{figure}
\begin{figure}[h]
  \centering
  \def\svgwidth{1\columnwidth} 
\executeiffilenewer{svg/strong3.svg}{strong3.pdf}%
{/usr/bin/inkscape -z -D --file="svg/strong3.svg" --export-pdf="strong3.pdf" --export-latex}
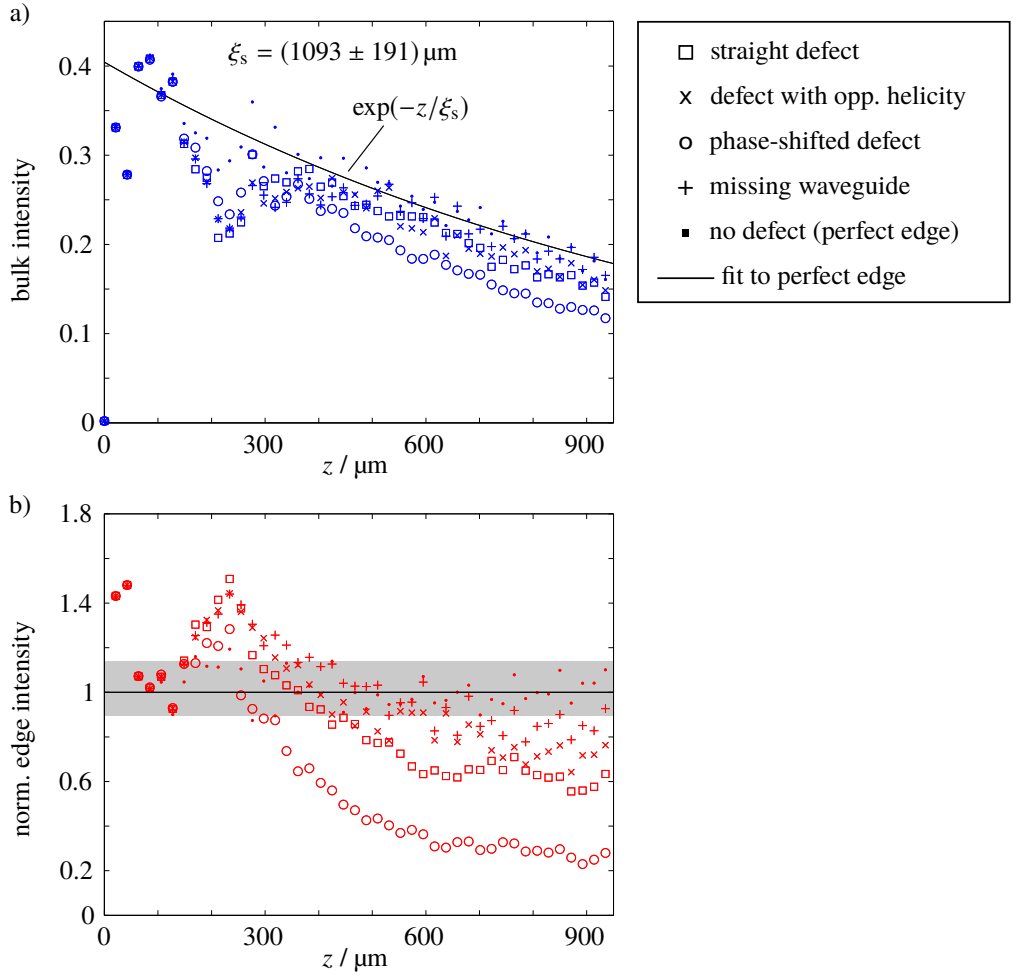%
  
  \caption{Numerical calculations for strong dynamic defects: Intensity in edge (red) and bulk (blue) for samples with and without defects. b) normalized to fitting curve for the edge intensity of the defect-free sample to remove overall losses. The dip in (a) and peak in (b) around $z\approx \SI{200}{\um}$ results from counting waveguides surrounding the defect as edge waveguides (see white circles in Figure \ref{fig:results_strong}).}
  \label{fig:opti_strong}
\end{figure}
In the simulation, three waveguides at the edge are excited by Gaussian beams with a phase difference of $\pi$. This corresponds to a transversal wavevector component of $k_x\sqrt{3}a=\pi$.
Videos showing the propagation of the edge mode around the defect were constructed via the simulations and can be found in the Supplementary \ref{supplementary}.
For weak defects the videos basically show that light mainly moves through the defect waveguide, in accordance to the measurements. However, the overall loss in intensity due to bending is severe. 
For strong dynamic defects the videos show another interesting effect, that is not captured by the single images obtained by measurements: Some defects seem to retard part of the initial wave-packet and split it in parts, that subsequently move along the edge separately. This is visible in the video for the straight defect and the defect with opposite helicity.

For quantitative analysis the intensity in the edge waveguides as well as in the bulk waveguides is summed up for each $z$-step and normalized to the intensity at the excitation point ($z=0$). For some samples with defects, the edge mode partially moves through and partially around the defect (see Videos and section \ref{results}). Therefore, also the waveguides immediately surrounding the defect are counted as edge waveguides in these cases. Waveguides considered as edge waveguides are indicated by white circles in Figure \ref{fig:results_strong}. Note, that the edge modes are localized with some finite localization length and thus extend into bulk waveguides. Therefore we make a small error in counting only the intensity in the edge waveguides. However, thoroughly distinguishing between intensity from edge and bulk modes in the same waveguide is not possible.

The simulated situation corresponds to a scattering experiment:
A wave-packet (excited at the three waveguides at $z=0$) travels along the edge, meets the defect and interacts with it for a certain time interval (\SI{100}{\um}< $z_{\mathrm{scatter}}$<\SI{600}{\um}), and then moves on along the edge. To determine if the defect causes scattering into the bulk, we examine the intensity in the edge and in the bulk waveguides. We need to look at both intensities, as we have to distinguish between two effects: scattering into the bulk, and losses into the continuum induced by the defect.
In addition to that, overall losses are present in our system. These are mainly absorption and bending losses and lead to an exponential decay of the intensities (compare for Figure \ref{fig:opti_weak} and \ref{fig:opti_strong} (a)). For the samples with large $a$ (weak dynamic defects, Figure \ref{fig:opti_weak}) the bending losses are bigger. We expected this since the helix pitch is smaller than for the other set of parameter and also $\lambda$ is increased. To separate these overall losses from the dynamics, we fit an exponential decay to the curves of the defect-free sample (perfect edge) for $z\gg z_{\mathrm{scatter}}$, where the influence of the defect is negligible. The edge intensities in Figure \ref{fig:opti_weak} and \ref{fig:opti_strong} (b) are then normalized to the respective fit.
The grey shading indicates the intensity range for the defect-free sample.  We assume that intensity, that is radiated off due to bending, is picked up by neighboring waveguides and thus leads to fluctuations of the data.

Figures \ref{fig:opti_weak} and \ref{fig:opti_strong} show two things: First, the exponential decay rate for $z\gg z_{\mathrm{scatter}}$ is not influenced by any of the examined defects, as one would expect. This means, that the normalized curves are approximately constant for 
$z\gg z_{\mathrm{scatter}}$.

Second, the defects do not lead to scattering into the bulk, but only to losses into the continuum.  Figure \ref{fig:opti_weak} and \ref{fig:opti_strong} (a) shows, that the intensity in the bulk does not rise above the value for the defect-free sample. The drop in the intensity in the edge therefore has to be interpreted as loss into the continuum. This drop is most prominent for the sample with the phase-shifted and with the straight strong dynamic defect (Figure \ref{fig:opti_strong} (b)). Considering both the bulk and edge intensity indicates, that rather than scattering into the bulk, the intensity is radiated away. This can be explained by mode mismatching, as the phase-shifted and the straight defect are overlapping with their neighbors at certain $z$. 

In conclusion, the numeric simulation indicates that dynamic defects do not lead to scattering into the bulk. Thus, the topological edge mode is still robust in their presence.

\section{Conclusion and Outlook}
The examined single dynamic defects seem to have no influence on the robustness of a chiral edge mode in a Floquet topological insulator. This is confirmed by measurements as well as numerical calculations. In all cases of the studied defects no scattering into the bulk occurs, rather the mode partially moves around the defect and partially through it. Even when the defect is overlapping with its neighbors at times, the edge mode is surprisingly robust against scattering into the bulk. In that case however, a lot of light is radiated off due to mode-mismatching. Further investigation is needed to see if a bigger amount of dynamic defects might lead to a different behavior (i.e. scattering into the bulk).

The method used to fabricate the waveguide samples is quite flexible. For example it allows to change the refractive index contrast easily by infiltrating the inverse sample with different materials.
In the same way, nonlinear materials can be used to form waveguides, to observe the effects of nonlinear waveguides on the topological robustness.

\section*{Funding}
F.L. is supported by a fellowship through the Excellence Initiative MAINZ (DFG/GSC 266).

\section*{Acknowledgments} 
We would like to acknowledge support by the Nano Structuring Center Kaiserslautern and by the Deutsche Forschungsgemeinschaft through CRC/Transregio 185 OSCAR.

\section*{Supplementary} \label{supplementary}
Videos, showing the propagation of the edge mode:

\noindent
https://www.dropbox.com/sh/8r4h02thqjvyuyu/AAA3Bxtz5yHzACooQxpIB2BHa?dl=0

\noindent
The videos were obtained by numerical simulation, using OptiBPM, which relies on the beam propagation method.
For better visibility, the colorbar indicating the intensity is decreasing exponentially with the propagation distance $z$, to account for bending losses. 

In the case of weak defects this leads to the impression that a large proportion of intensity is in the bulk. Note however, that the maximal intensity scaling for large $z$ is very low ($\approx 3\cdot10^{-3}$).
The intensity in the bulk results from an overlap of the excited modes with bulk modes, as the transversal wavevector component is not sharp, but distributed around $k_x\sqrt{3}a=\pi$ to some amount (indicated by gray region in the band structure, Figure \ref{fig:bandstructure} (a)). 
Those bulk modes (initially generated at the edge) propagate to the inner of the sample, whereas the proper edge modes stay at the edge. This is why the intensity in the bulk increases in about the first $\SI{100}{\um}$ of propagation (see Figure \ref{fig:opti_weak} (a) and \ref{fig:opti_strong} (a)), even for the sample with the perfect edge.

\bibliography{literatur}
\end{document}